\documentclass[aip, preprint]{revtex4-1}
\usepackage{amsmath}    
\usepackage{graphicx}   
\usepackage{textcomp}

\usepackage{hyperref, color}
\definecolor{linkColor}{rgb}{0.8,0,0}
\definecolor{darkred}{rgb}{0.8,0,0}
\hypersetup{pdfborder={0 0 0},colorlinks=true,urlcolor=linkColor,citecolor=linkColor}

\begin{document}

\title{Percolation of Ion-Irradiation-Induced Disorder in Complex Oxide Interfaces}

\author{Bethany Matthews}
\affiliation{Energy and Environment Directorate, Pacific Northwest National Laboratory, Richland, Washington 99352}

\author{Michel Sassi}
\affiliation{Physical and Computational Sciences Directorate, Pacific Northwest National Laboratory, Richland, Washington 99352}

\author{Christopher M. Barr}
\affiliation{Center for Integrated Nanotechnologies, Sandia National Laboratories, Albuquerque, New Mexico 87195}

\author{Colin Ophus}
\affiliation{NCEM, Molecular Foundry, Lawrence Berkeley National Laboratory, Berkeley, California 94720}

\author{Tiffany Kaspar}
\affiliation{Physical and Computational Sciences Directorate, Pacific Northwest National Laboratory, Richland, Washington 99352}

\author{Weilin Jiang}
\affiliation{Energy and Environment Directorate, Pacific Northwest National Laboratory, Richland, Washington 99352}

\author{Khalid Hattar}
\affiliation{Center for Integrated Nanotechnologies, Sandia National Laboratories, Albuquerque, New Mexico 87195}

\author{Steven R. Spurgeon}
\email{steven.spurgeon@pnnl.gov}
\affiliation{Energy and Environment Directorate, Pacific Northwest National Laboratory, Richland, Washington 99352}

\date{\today}

\begin{abstract}

Mastery of order-disorder processes in highly non-equilibrium nanostructured oxides has significant implications for the development of emerging energy technologies. However, we are presently limited in our ability to quantify and harness these processes at high spatial, chemical, and temporal resolution, particularly in extreme environments. Here we describe the percolation of disorder at the model oxide interface LaMnO$_3$ / SrTiO$_3$, which we visualize during \textit{in situ} ion irradiation in the transmission electron microscope. We observe the formation of a network of disorder during the initial stages of ion irradiation and track the global progression of the system to full disorder. We couple these measurements with detailed structural and chemical probes, examining possible underlying defect mechanisms responsible for this unique percolative behavior.

\end{abstract}

\maketitle

\section{Introduction}

Interface engineering of oxide thin films is one of the great achievements of materials science, underpinning exotic physics\cite{Hwang2012} and giving rise to advanced computing and energy technologies.\cite{Martin2017} Because of strong interplay among lattice, charge, and spin degrees of freedom, even slight fluctuations in local order at oxide interfaces can greatly impact behaviors such as electronic\cite{Huang2018} and ionic conductivity.\cite{Fabbri2010} Order can be disrupted via numerous defect mechanisms, such as local structural perturbations\cite{Ismail-Beigi2017, Shamblin2016a} or the formation of cation and anion vacancies.\cite{Gunkel2020,Zhang2015} The desire to understand these mechanisms has motivated efforts to map defect formation pathways during oxide synthesis\cite{MacManus-Driscoll2020,Brahlek2018} and exposure to extreme environments.\cite{Spurgeon2020b,Zhang2018a,Beyerlein2013} Understanding oxides in extremes of temperature and irradiation is particularly important, since devices such as solid oxide fuel cells (SOFCs),\cite{Shamblin2016a,Tuller2011} durable nuclear waste forms,\cite{Ewing2004,Sickafus2000} and space-based electronics\cite{Cramer2016} must deliver reliable long-term performance under challenging conditions. At a more fundamental level, the nature of order-disorder processes represents a grand scientific question, with implications for the design of high entropy alloys\cite{George2019} and ceramics,\cite{Oses2020} strongly correlated quantum systems,\cite{Vepsalainen2020a} and more.

To dictate order-disorder behavior, we must be able to visualize and direct defect formation processes at high spatial, chemical, and temporal resolution. The community has a long, successful history of using tailored ion irradiation to manipulate and induce defect populations in bulk metals\cite{Was2015,Odette2008} and oxides.\cite{Meldrum1998,Lumpkin2009,Sickafus2000,Shamblin2016,Lang2010} While recent studies\cite{Zhang2018a,Martinez2016,Beyerlein2015,Beyerlein2013} have showcased the promising properties of nanostructured systems, such as enhanced defect annihilation and radiation hardness, past work has focused almost exclusively on metals and bulk oxides.\cite{Zhang2018a} A handful of studies of model oxide interfaces have combined controlled ion irradiation and local characterization tools to identify unique behaviors such as anti-site defect buildup,\cite{Kreller2019} oxygen vacancy formation,\cite{Spurgeon2020} and orientation-/chemistry-dependent amorphization behavior.\cite{Kaspar2017,Aguiar2014,Aguiar2014a} These studies show that there is an interplay between defect formation energy, which depends on the interface configuration (e.g. strain, charge state, chemistry)\cite{Spurgeon2020,Dholabhai2014,Aguiar2014,Aguiar2014a,Zhuo2011} and defect kinetics, particularly defect mobility in different interface components.\cite{Zhuo2012a} However, most prior work has focused on static snapshots of these materials and has not possessed sufficient spatiotemporal resolution to probe the kinetics of local defect formation, particularly during the initial loss of crystallinity in oxide thin films.

Moving beyond traditional static characterization approaches toward \textit{in situ} methods will allow us to capture the initial onset and evolution of local defects. \textit{In situ} (scanning) transmission electron microscopy ((S)TEM) has been widely employed by the radiation effects community to examine local defects in materials,\cite{Parrish2021,Lian2009,Birtcher2005} providing an invaluable window into irradiation-induced disorder, particularly in the low dose regime. These methods have long been used to examine bulk ceramics,\cite{Ye2011, Lian2009, Zhang2005a,Wang1998,Zinkle1992} revealing radiation-induced defect formation pathways and kinetics. Local (S)TEM probes are particularly well-suited to examining nanostructured oxide interfaces,\cite{Spurgeon2017a} whose non-equilibrium behavior can deviate greatly from bulk materials, but surprisingly little work has been done on complex oxide thin film systems in the context of radiation damage. More broadly, the microscopy community has also recognized the need for new data-driven approaches to characterization of transient processes, needed to detect and quantify salient features during high-speed imaging.\cite{Hattar2021,Spurgeon2020c,Taheri2016}

Here we visualize the evolution of local disorder at a LaMnO$_3$ (LMO) / SrTiO$_3$ (STO) (001) perovskite oxide interface using ion irradiation coupled with \textit{in situ} high-resolution transmission electron microscopy (HRTEM) imaging at the I$^3$TEM irradiation facility at Sandia National Laboratories. We observe the initial onset of disorder and track its progression over continued irradiation. Using a Fourier filtering time series approach, we show that initial radiation damage is accommodated by the percolation of amorphous regions throughout the crystalline LMO matrix. In addition, we find evidence for a preserved crystalline interface region even at the highest fluence studied, whose origin we examine in the context of electronic structure calculations and past work. Taken together, these results demonstrate the power of high-resolution \textit{in situ} approaches to derive complex disordering pathways at oxide interfaces. More broadly, the data processing approaches we demonstrate may be applied to other dynamic studies of complex materials phase transitions.

\section{Results and Discussion}

\begin{figure*}
\includegraphics[width=0.62\textwidth]{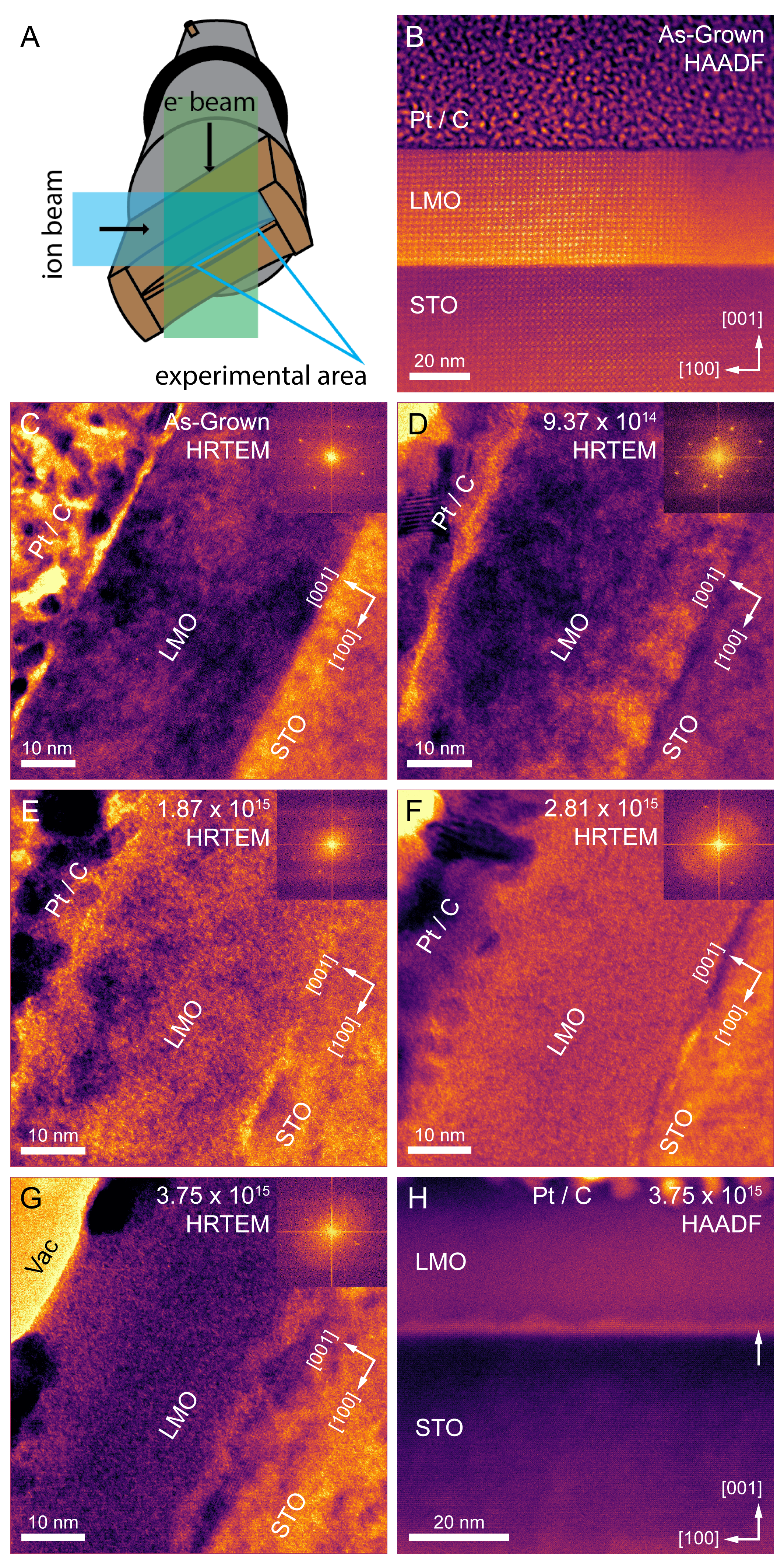}
\caption{Evolution of local disorder with fluence. (A) Schematic of the irradiation geometry. (B--C) Colorized cross-sectional STEM-HAADF and HRTEM images of the as-grown film. (D--G) HRTEM images for fluences of $9.37 \times 10^{14}$, $1.87 \times 10^{15}$, $2.81 \times 10^{15}$, $3.75 \times 10^{15}$ Au$^{4+}$ cm$^{-2}$, respectively. (H) STEM-HAADF image of the $3.75 \times 10^{15}$ Au$^{4+}$ cm$^{-2}$ fluence sample. (Higher-resolution images are provided in Supporting Information Note 5). \label{ex_situ}}
\end{figure*}

We first consider the overall degradation of the system from the crystalline, as-grown condition to its disordered state. Figure \ref{ex_situ}.A shows the configuration of the electron and ion beams during irradiation, with the sample inclined 30$^{\circ}$ in the X tilt direction to minimize shadowing of the ion beam. Figure \ref{ex_situ}.B shows a cross-sectional STEM-HAADF image of the starting epitaxial 40 nm LMO film along the STO [100] crystallographic zone-axis prior to irradiation. In this mode, the directly-interpretable atomic number ($Z^{\sim1.7}$) contrast reveals a clear difference between the film and substrate, with minimal intermixing, excellent epitaxy, and no extended defects. Figure \ref{ex_situ}.C shows a HRTEM image of the same film in the I$^3$TEM system, with the interface rotated approximately 60$^{\circ}$ about the [100] zone (the normal direction to the image plane). Phase contrast in HRTEM is less directly interpretable than STEM-HAADF, but the inset fast Fourier transform (FFT) and clear lattice fringes confirm the crystalline starting condition. From this point, a series of four irradiations were performed in 1 hour increments with 2.8 MeV Au$^{4+}$ ions, as shown in Figures \ref{ex_situ}.D--G, with the sample inclined during irradiation and then tilted back to the zone for imaging. Each irradiation step corresponds to a fluence of $9.37 \times 10^{14}$ Au$^{4+}$ cm$^{-2}$. These figures show a gradual amorphization sequence, starting from the LMO film and extending to the STO substrate. A careful inspection of the images and associated FFTs reveals the emergence of local amorphous patches in the LMO at a fluence of $9.37 \times 10^{14}$ Au$^{4+}$ cm$^{-2}$ (Figure \ref{ex_situ}.D), with less apparent damage in the STO film. By a fluence of $1.87 \times 10^{15}$ Au$^{4+}$ cm$^{-2}$ (Figure \ref{ex_situ}.E), the patches have grown more extensive in the LMO, and the STO side also exhibits obvious disorder; this is accompanied by a decrease in the intensity of the FFT reflections. At a fluence of $2.81 \times 10^{15}$ Au$^{4+}$ cm$^{-2}$ (Figure \ref{ex_situ}.F), the LMO film is almost entirely amorphous, except for a thin, several nm-thick band at the interface. Finally, at a fluence of $3.75 \times 10^{15}$ Au$^{4+}$ cm$^{-2}$ (Figure \ref{ex_situ}.G), the LMO film appears similarly disordered to its state at a fluence of $2.81 \times 10^{15}$ Au$^{4+}$ cm$^{-2}$, but it is clearly thinner in the beam direction (as indicated by the larger vacuum region at the top left corner of the image), likely due to sputtering of the film under the ion bombardment. In addition, the STO layer is extensively damaged and the only remaining lattice fringes are from a 2--3 nm band at the film-substrate interface. There is almost no periodic signal left in the FFT, reflecting the loss of overall crystalline order. STEM-HAADF images of the $3.75 \times 10^{15}$ Au$^{4+}$ cm$^{-2}$ condition (Figure \ref{ex_situ}.H) show that some lattice fringes are still present in the bulk of the STO, but the bulk of the LMO is completely amorphous. Nonetheless, there is a pronounced and persistent crystalline band on the film side of the interface (marked by the arrow). We note that SRIM simulations, shown in Supporting Information Note 3, indicate a trace amount of Au ions may be retained in the TEM foil, which may partly influence the kinetics of the radiation response. However, a similar irradiation response has been observed in the La$_2$Ti$_2$O$_7$ / STO system,\cite{Spurgeon2020} where a 5--10 nm crystalline band persisted on the STO side of the interface. While the origins and mechanisms leading to the resistant interface behavior are yet not fully understood, we do note the presence of some cation intermixing in the irradiated sample (shown in Supporting Information Note 4), which may affect its stability. Overall, this points to the broader trend for perovskite oxide interface systems to remain crystalline under irradiation environments.

\begin{figure*}
\includegraphics[width=\textwidth]{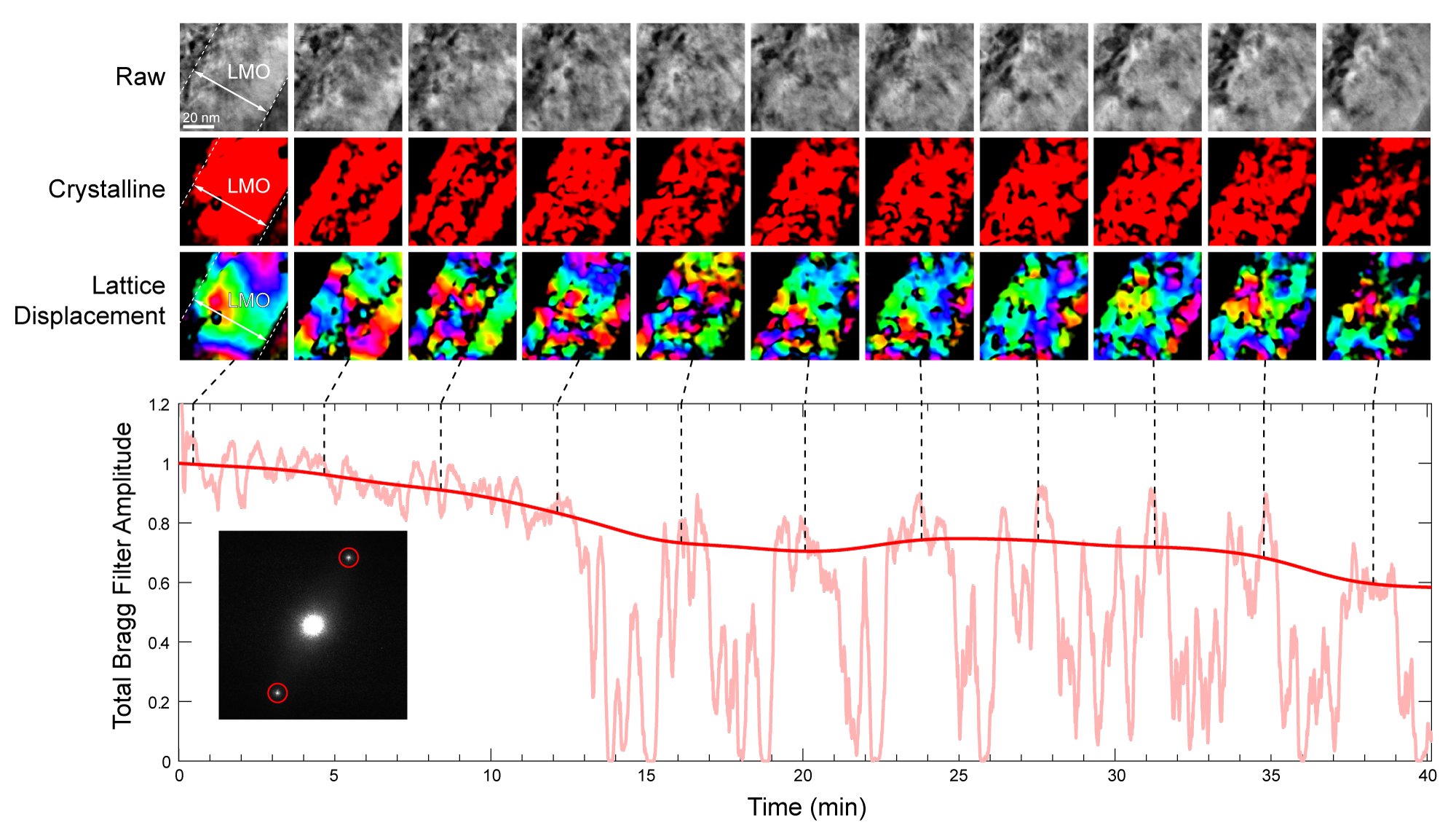}
\caption{Percolation of initial local lattice disorder. Top to bottom: Raw frames, crystallinity maps, displacement maps, and total Bragg filter amplitude as a function of frame time. The dark red line corresponds to an iteratively fitted time-average of the amplitude data and the inset shows the masked FFT reflections used for the analysis. Total combined fluence over 40 minutes is $6.25 \times 10^{14}$ Au$^{4+}$ cm$^{-2}$. \label{in_situ}}
\end{figure*}

We focus next on the initial stages of the irradiation process, probing the initial percolation of disorder and its buildup to the more global amorphization shown in Figure \ref{ex_situ}.H. For this experiment, we captured a high-speed time series with the sample inclined at 30$^{\circ}$ in the X tilt direction to maximize exposure to the ion beam. At this high tilt angle it is no longer possible to directly resolve atomic columns in the [100] zone-axis, and we instead image (100)-type lattice planes. However, the remarkable stability of the sample during the extended period of irradiation (40 minutes, total fluence $6.25 \times 10^{14}$ Au$^{4+}$ cm$^{-2}$) provided a unique window into the initial stages of disorder. We employ a temporal Fourier filtering approach, analogous to time-resolved geometric phase analysis (GPA),\cite{Hytch1998} processing the raw movie frames into maps of local crystallinity and lattice displacement. In addition, we quantify the total Bragg filter amplitude, which effectively measures the spatial abundance of the selected (100)-type lattice domains in any given frame. It is important to note that while the overall sample was quite stable, these measurements are very challenging and some periodic bending/rotation of the sample due to local heating did occur; since this results in large random drops in the Bragg filter amplitude, a trend line was fit to the Bragg filter amplitude only where crystalline signal is present (see Supporting Information Methods for details). These measurements are shown in Figure \ref{in_situ} and Supporting Information Movie S1.

We see that there is initially a large contiguous block of crystalline region, as expected. By 5 minutes, some disorder emerges in the film, beginning at its center and extending laterally, but the Bragg amplitude is still around 95\% of the starting condition. Disorder is accompanied by significant local lattice displacement around the defective regions. At this stage, misfit dislocations emerge that run from the lower left to upper right corner of the frames. Between 9 and 19 minutes, the percolation extends from the film center to its surface and the substrate interface. At this point there is a substantial drop in the Bragg amplitude to nearly 80\%, accompanied by a still more complex pattern of lattice rotation and crystalline domains, some $< 10$ nm in size. The misfit dislocations appear to reconnect, forming repaired epitaxial domains, but there is a clear increase in disorder. Between 19 to 33 minutes the overall distribution remains steady near 80\%, but between 33 and 37 minutes the percolation continues and the crystalline domains become increasingly sparse and disconnected. In the final condition around 40 minutes, only about 60\% of the film is still similar to the starting crystalline condition and there is a large disconnected network of ordered regions. These results point to the highly non-uniform evolution of the interface during the initial stages of irradiation.

\begin{figure*}
\includegraphics[width=\textwidth]{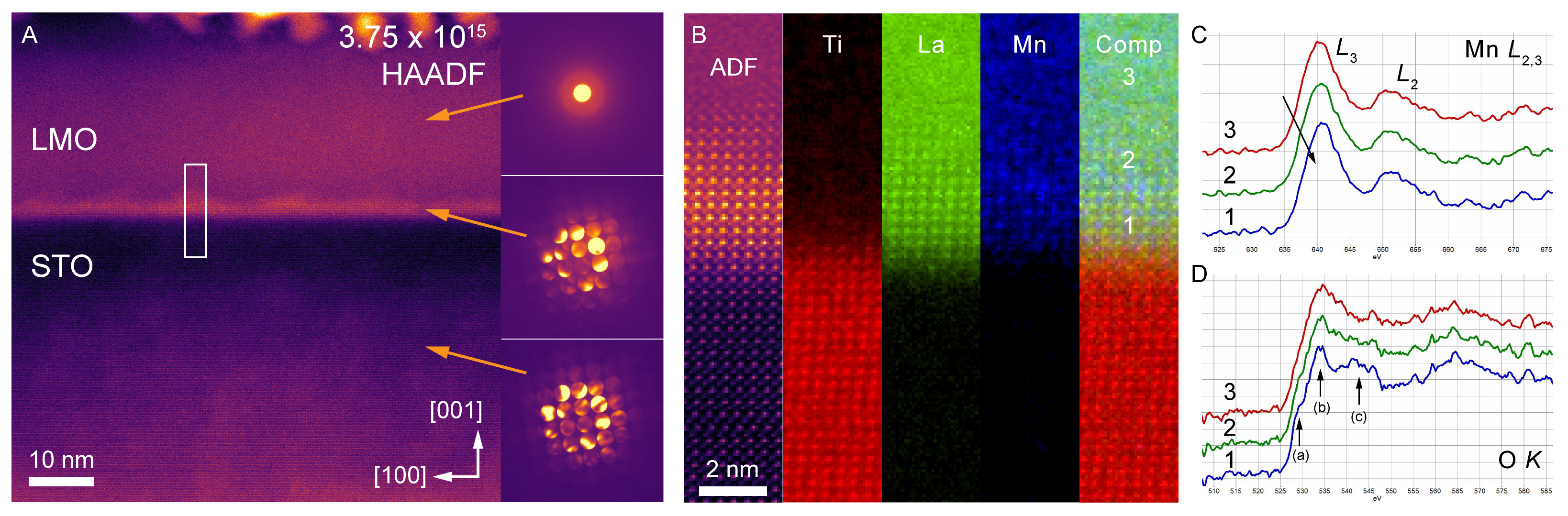}
\caption{Detail of the interface after irradiation. (A) Colorized cross-sectional STEM-HAADF image, with local PACBED patterns from the film, interface, and substrate, taken from the $3.75 \times 10^{15}$ Au$^{4+}$ cm$^{-2}$ fluence sample. (B) Colorized STEM-ADF and corresponding STEM-EELS Ti $L_{2,3}$, La $M_{4,5}$, Mn $L_{2,3}$, and composite maps taken from near the boxed region in (A). (C--D) Corresponding Mn $L_{2,3}$ and O $K$ spectra, respectively, from regions 1--3.  \label{eels}}
\end{figure*}

To further explore this non-uniform interface response in later stages of irradiation, we have performed local position-averaged convergent beam electron diffraction (PACBED) and electron energy loss spectroscopy (STEM-EELS) measurements. The former is a mode of scanning nanodiffraction\cite{Ophus2019} that can quantify crystallinity at the nanoscale and has been used before to examine local disorder in oxides,\cite{Savitzky2020,Spurgeon2020,Janish2019} while the latter has been used to examine chemical states in both pristine and irradiated oxides.\cite{Spurgeon2020b} As shown in Figure \ref{eels}.A, at a fluence of $3.75 \times 10^{15}$ Au$^{4+}$ cm$^{-2}$ the bulk of the LMO film has amorphized, resulting in a diffuse, ring-like PACBED pattern. In contrast, PACBED from the uniform $\sim3$ nm interface band exhibits strong diffraction disks; there is also evidence for more weakly diffracting patches that extend further into the film. Similarly, the STO side of the interface also exhibits strong diffraction disks, despite some distributed damage, consistent with past observations.\cite{Zhang2005a,Jiang2012a} Correlative STEM-EELS chemical analysis of the interface, shown in Figure \ref{eels}.B, reveals the gradual breakup of the crystalline film upon transitioning from the interface into the bulk, resulting in three distinct regions (labeled 1--3). Region (1) consists of a $\sim 3$ nm region of LMO that is strongly diffracting and highly chemically ordered on both the La and Mn sublattices. EELS spectra from this region contain a characteristic pre-peak (a), main peak (b), and secondary peak (c) in the O $K$ edge fine structure, as well as a clear Mn $L_{2,3}$ white-line doublet, consistent with previous observations.\cite{Kaspar2019a,Varela2009} Moving away from the interface and into the LMO, we move into a more weakly diffracting $\sim2$ nm Region (2) consisting of poorly defined perovskite atomic order. This region exhibits a reduction in the O $K$ pre-peak feature and loss of definition between the main and secondary peaks. In addition, there is a clear 0.5 eV shift of the Mn edge to lower energy loss, pointing toward underlying reduction. This region is followed by third and final amorphous region (3) with no extended ordering of La and Mn; now the O $K$ edge pre-peak has disappeared and the main edge features have blurred into one from the increased randomization of the sublattice. There is a further 0.5 eV Mn chemical shift, reflecting further Mn reduction. The origin of this reduction is likely the loss of oxygen (oxygen vacancy formation) and electronic reconfiguration during the ion irradiation process.\cite{Aguiar2014,Spurgeon2020}

In order to understand the energetics of oxygen vacancies at the LMO / STO interface, density functional theory (DFT) calculations have been performed, in which an oxygen vacancy has been introduced at different locations of the LMO / STO interface. These simulations used three different model interface terminations (i.e., two LMO / STO (001) and one LMO / STO (110)), as shown in Supporting Information Note 2. The two (100) interface models were built from the STO (100) substrate orientation, terminated either by SrO or TiO$_2$, yielding LMO / STO interface configurations with cation rows ordering as (Ti--Sr--Mn--La) and (Sr--Ti--La--Mn), respectively. As shown by the relative energy plots in Supporting Information Note 2, forming an oxygen vacancy on the LMO side of the interface is found more favorable than on the STO side. This behavior is likely due to the ability of Mn$^{4+}$ ions to reduce to Mn$^{3+}$ more readily than for Ti$^{4+}$ to reduce to Ti$^{3+}$. Supporting Information Note 2 also shows that creating an oxygen vacancy at the interface of the Ti-terminated LMO (100) / STO (100) configuration is less favorable than creating it at the Sr-terminated configuration. While the Ti-terminated interface provides a less favorable environment for oxygen vacancy creation compared to the Sr-terminated interface, the formation energy is still similar to that in STO. These results suggest that, in the LMO / STO system, the mechanisms for interface stabilization are not dominated by the oxygen vacancy creation. Nevertheless, these results overall show a net energy gain of 2 eV for vacancy formation on the LMO side of the interface relative to the STO side. This finding is in excellent agreement with the overall trend in the amorphization sequence that we observe experimentally. However, unlike in the case of LTO / STO,\cite{Spurgeon2020} no interfacial oxygen vacancy differences were observed to explain the retention of crystallinity at the interface, so the specific characteristics of the damage mechanism are likely different.

\section{Conclusions}

Using an \textit{in situ} imaging approach, we reveal the percolation of disorder in oxide thin film interfaces. Our results indicate the formation of a complex network of amorphization during the initial stages of irradiation, which progresses to global disorder over longer time scales. However, we also observe the preservation of a distinct crystalline interface region on the film side of the LMO / STO interface. Our calculations demonstrate a propensity for defect accumulation in the bulk of the LMO, which likely influences the course of the disordering processes. These results support the general trend toward retention of interfacial crystallinity, which appears to be unique to epitaxial oxides. Collectively, these results demonstrate the untapped potential of high-resolution spatiotemporal probes to survey the complex landscape of disorder and underscore the important role of interface engineering in mediating disordering processes.

\section{Acknowledgements}

This research was supported by the Nuclear Processing Science Initiative (NPSI) Laboratory Directed Research and Development (LDRD) at Pacific Northwest National Laboratory (PNNL). PNNL is a multiprogram national laboratory operated for the U.S. Department of Energy (DOE) by Battelle Memorial Institute under Contract No. DE-AC05-76RL0-1830. C.O. acknowledges support from the DOE Early Career Research Program. The STEM imaging shown was performed in the Radiological Microscopy Suite (RMS), located in the Radiochemical Processing Laboratory (RPL) at PNNL. Sample preparation was performed at the Environmental Molecular Sciences Laboratory (EMSL), a national scientific user facility sponsored by the Department of Energy's Office of Biological and Environmental Research and located at PNNL. \textit{In situ} ion irradiation work was performed at the Center for Integrated Nanotechnologies, an Office of Science User Facility operated for the U.S. DOE. Sandia National Laboratories is a multimission laboratory managed and operated by National Technology \& Engineering Solutions of Sandia, LLC, a wholly owned subsidiary of Honeywell International, Inc., for the U.S. DOE's National Nuclear Security Administration under contract DE-NA-0003525. Work at the Molecular Foundry was supported by the Office of Science, Office of Basic Energy Sciences, of the US DOE under Contract DE-AC02-05CH11231. The views expressed in the article do not necessarily represent the views of the U.S. DOE or the United States Government.

\section{Competing Interests Statement}

The authors declare no competing interests.

\section{Data Availability Statement}

The data used in this study are available from the authors upon reasonable request.

\section{Supporting Information}

The Supporting Information is available free of charge at X. Details of methods used, DFT calculations, SRIM calculations, sequence of the irradiation, and a movie of the irradiation are available.

\section{Table of Contents Figure}

\begin{figure*}[h]
\includegraphics[width=0.7\textwidth]{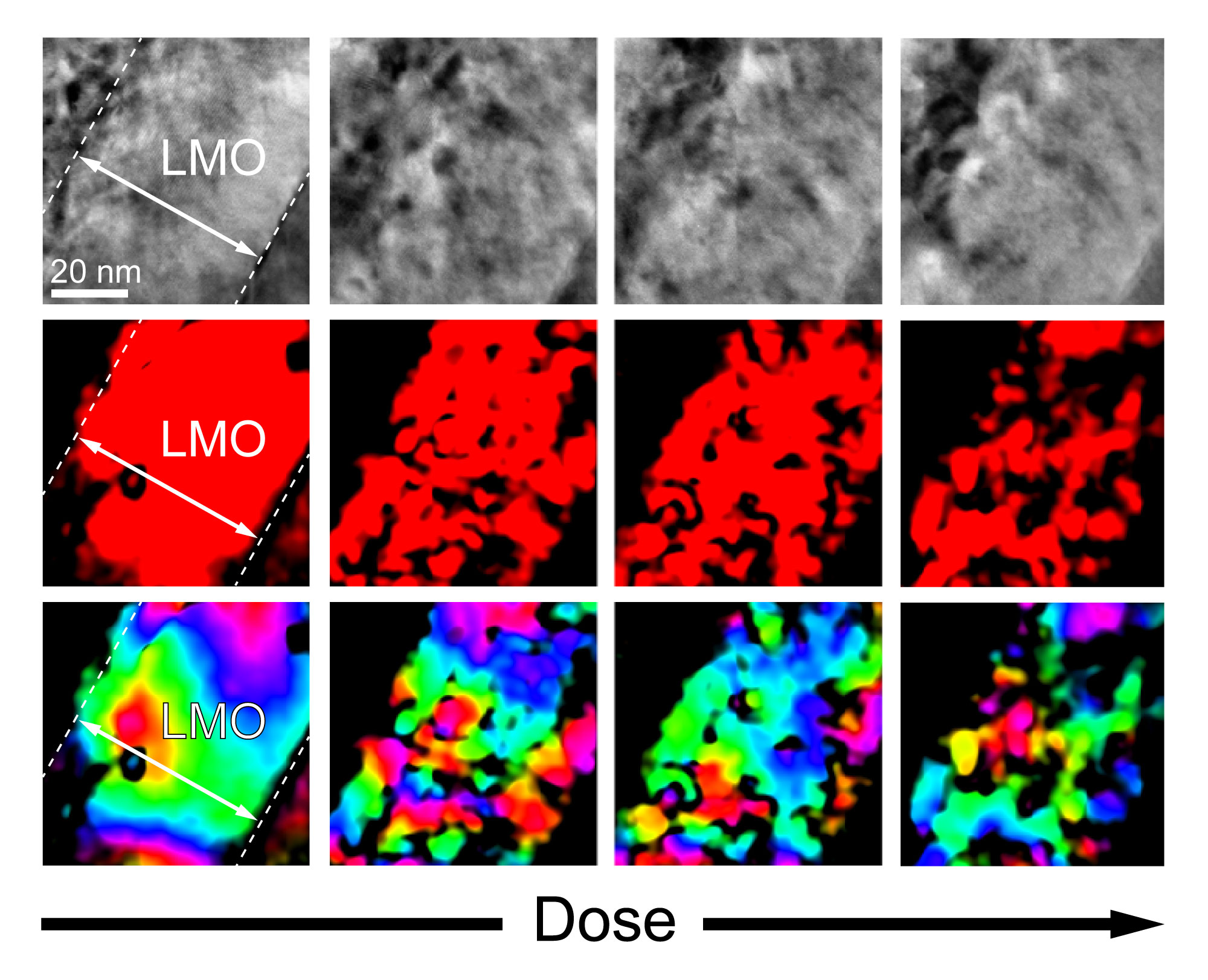}
\end{figure*}

\clearpage

\bibliography{references}

\end{document}


\section*{Supporting Information Note 1: Methods}

\subsection{Thin Film Growth}

Nominally 40 nm-thick epitaxial, single-crystal LaMnO$_3$ films were deposited on SrTiO$_3$ (001) substrates by oxygen-plasma-assisted molecular beam epitaxy (OPA-MBE), as described previously.\cite{Kaspar2019a} Briefly, deposition occurred at 625 $^{\circ}$C with oxygen supplied from a differentially pumped electron cyclotron resonance (ECR) microwave plasma source at a chamber pressure of 2.4 × 10$^{-6}$ Torr. La and Mn were supplied from effusion cells to realize an LMO growth rate of 45 s u.c.$\rm{}^{-1}$ ($\sim0.85$~\AA \, s$^{-1}$). After deposition, the films were cooled in the flow of activated oxygen from the ECR plasma source.

\subsection{Transmission Electron Microscopy}

Cross-sectional TEM samples were prepared using a FEI Helios NanoLab DualBeam Focused Ion Beam (FIB) microscope and a standard lift out procedure. Samples were mounted into a slot at the end of a lift out half grid to accommodate the ion beam geometry. We note that a slight wedge shape is inevitable, but the sample is nonetheless quite flat. High-angle annular dark field (STEM-HAADF) images of the as-grown and post-irradiated samples were collected using a probe-corrected JEOL GrandARM-300F STEM operating at 300 kV with a semi-convergence angle of 27.5 mrad and a collection angle of 82--186 mrad. Position-averaged convergent beam electron diffraction (PACBED) patterns were collected with a reduced convergence semi-angle of $\alpha \approx 4$ mrad to reduce diffraction disk overlap. STEM electron energy loss spectroscopy (STEM-EELS) measurements were conducted in DualEELS mode, with a at a 0.25 eV ch$^{-1}$ dispersion yielding an approximate $\sim0.75$ eV energy resolution. The spectra were subsequently corrected for zero loss energy drift, but no denoising was applied. The maps shown in the supplement were collected with $4 \times$ energy binning to improve signal-to-noise.

\textit{In situ} ion irradiation was performed using the I$^3$TEM system at CINT, which consists of a JEOL 2100 LaB$_6$ TEM modified to incorporate an ion beam.\cite{Hattar2014} We utilized 2.8 MeV Au$^{4+}$ ions with a current density of $\sim$1.6 nA cm$^{-2}$. On-axis high-resolution transmission electron microscopy (HRTEM) imaging was performed prior to irradiation along the STO [100] zone-axis and the sample was subsequently inclined at 30$^{\circ}$ in the X tilt direction to minimize shadowing of the ion beam during each irradiation step. Irradiations were performed in 1 hour steps, corresponding to a fluence of $9.37 \times 10^{14}$ Au$^{4+}$ cm$^{-2}$ per step, with the electron beam on and illuminating the sample. We note that prior work\cite{Lian2009} has shown that electron beam illumination can influence the radiation response of the material, but we did not systematically explore this effect. After every hour, the sample was tilted back onto the STO [100] zone and additional HRTEM imaging was performed. This process was repeated for a total irradiation time of 4 hours, yielding a total fluence of $3.75 \times 10^{15}$ Au$^{4+}$ cm$^{-2}$. Movies of each irradiation step were taken at $1024 \times 1024$ px$^2$ resolution with a 3.56 frames per second framerate.

\subsection{Post-Processing of \textit{In Situ} Data}

Analysis of the \emph{in situ} HRTEM time series experiment was performed with custom Matlab codes. We used a similar filtering method to the Bragg filtering used for geometric phase analysis (GPA) in \citet{Hytch1998}. First, we measured the position of the lone Bragg peak pair visible throughout the time series, and computed a Bragg filter mask centered on one of these peaks using a 2D Gaussian function with a standard deviation equal to 0.0025 1 px$^{-1}$. The location of these Bragg peaks is shown inset in main text Figure 2 as red circles. Next, for each image in the time series, we computed the 2D Fourier transform of the image, multiplied it by the Bragg filter, and then took the inverse 2D Fourier transform.

The resulting complex image was used in two ways; first, the amplitude was used to create a map of the spatial distribution where crystalline signal was present, shown for some images in the second row of main text Figure 2. The output minimum and maximum were scaled be equal to 0.004 and 0.008 of the normalized image intensity respectively. This was done to eliminate false positive signals, and to saturate the output value to 1 where crystalline signal was strongly visible. 

Next, we also used the phase of the complex output to map the lattice displacement, by dividing out the plane wave defined by the Bragg vector $\exp{ -2 \pi i \, {\bf q} \cdot {\bf r} }$, where ${\bf q}$ is the 2D Bragg vector and ${\bf r}$ is the spatial coordinate vector.\cite{Hytch1998} The resulting phase image divided by $2 \pi$ defines the local shift of the lattice, which we have plotted in the hue image channel in the third row of Figure 2. The amplitude mask was also applied to this row of images, to avoid plotting the lattice displacement in regions of pure noise (no crystalline signal present). The resulting hue images can show the present of dislocations and other lattice disconnections wherever the color field is discontinuous.

Finally, the total 2D amplitude image signal normalized to 1 at time 0 is plotted at the bottom of main text Figure 2. A best-fit trend line is also plotted, which was iteratively fit to values $\geq 75\%$ of the local trend line value, and smoothed with a 401 frame long Gaussian distribution. The resulting trend line tracks the mean Bragg filter amplitude in the frames where the sample is on zone and crystalline signal is present.

\subsection{SRIM Simulations}

The design of the \textit{in situ} Au ion irradiation experiments was accomplished based on SRIM (Stopping and Range of Ions in Matter) simulations,\cite{Ziegler2010} where the specific gravities of LaMnO$_3$ and SrTiO$_3$ used are 6.52 and 5.12 g cm$^{-3}$, respectively. The full damage cascade mode was used in the simulation, where the threshold displacement energies of $E_d$ (La) = 80 eV, $E_d$ (Mn) = 70 eV and $E_d$ (O) = 45 eV in LaMnO$_3$ and $E_d$ (Sr) = 80 eV, $E_d$ (Ti) = 70 eV and $E_d$ (O) = 45 eV in SrTiO$_3$ were assumed.\cite{Zhang2008a} 2.8 MeV Au ions at an incidence angle of $60^{\circ}$ were selected for irradiation to provide a balanced combination of dose rate and retained Au percentage in a TEM foil estimated to be $\sim80$ nm in thickness. These data are described in Supporting Information Note 3.

\subsection{Density Functional Theory}

The relative formation energy of oxygen vacancy has been calculated within the density functional theory (DFT) framework, as implemented in the VASP code.\cite{Kresse1996} All the simulations used the PBEsol exchange correlation functional,\cite{Perdew2008} spin-polarization, and a $4 \times 4 \times 1$ Monkhorst-Pack\cite{Monkhorst1976} $k$-point mesh to sample the Brillouin zone. The cutoff energy for the projector augmented wave\cite{Blochl1994} pseudo-potential was 500 eV, with convergence criteria of $10^{-5}$ eV per cell for the energy and $10^{-3}$ eV \AA$^{-1}$ for the force components.

Computational models of the LMO / STO interface used a periodic symmetric slab, such that LMO was sandwiched on both sides of a 4.5 unit cell of STO slab (i.e., LMO / STO / LMO). The lattice parameters were fixed and used the experimental lattice parameter for STO ($a = 3.905$~\AA). The simulation of the LMO / STO (001) interface used lateral lattice parameters of $a \sqrt{2} \times a \sqrt{2}$ and 76~\AA \, in the out-of-plane direction to avoid interactions between periodic images. Similarly, the simulation of the LMO / STO (110) interface used lateral lattice parameters of $a \sqrt{2} \times a \sqrt{2}$ and 93~\AA \, in the out-of-plane direction. In order to avoid artificial dipole effects, oxygen vacancies were symmetrically created on both sides from the center of the STO slab.

\clearpage

\section*{Supporting Information Note 2: Energy Cost of Oxygen Vacancy Formation}

\begin{figure*}[h]
\includegraphics[width=\textwidth]{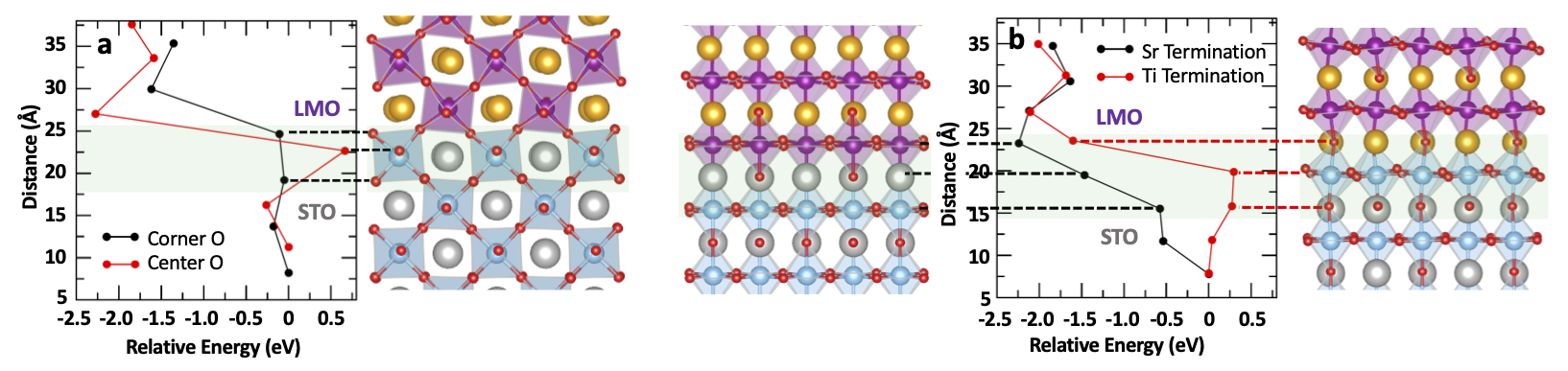}
\caption{Relative energy of oxygen vacancy creation across the LMO / STO interface for the (110) (a) and (001) (b) substrate orientations. For the (001) orientation, the Sr- and Ti-terminations are shown in left and right ball and stick models respectively. The La atoms are represented in yellow, O atoms in red, Sr atoms in grey, Mn atoms in purple, and Ti atoms in light blue. \label{si_dft}}
\end{figure*}

\clearpage

\section*{Supporting Information Note 3: SRIM Simulations of Ion Damage}

\begin{figure*}[h]
\includegraphics[width=\textwidth]{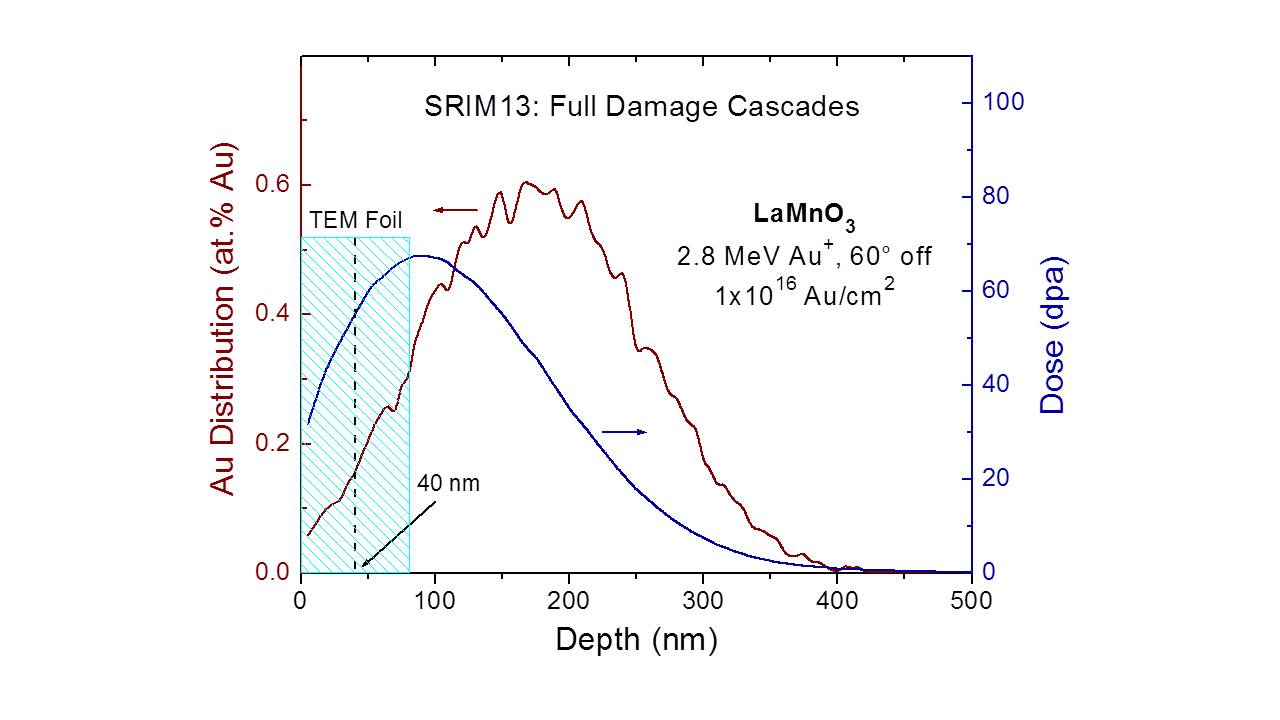}
\caption{SRIM full damage cascade calculations for LaMnO$_3$. \label{si_srim_lmo}}
\end{figure*}

\begin{figure*}[h]
\includegraphics[width=\textwidth]{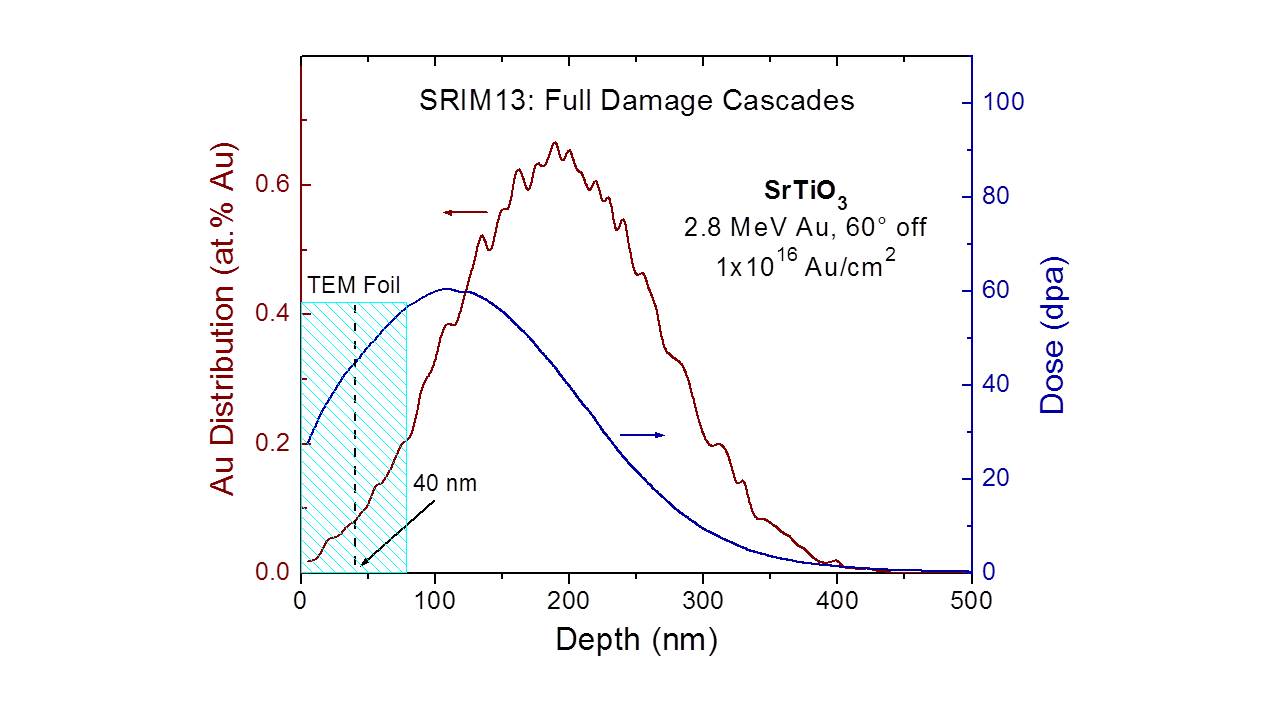}
\caption{SRIM full damage cascade calculations for SrTiO$_3$. \label{si_srim_sto}}
\end{figure*}

SRIM simulations were performed for an ion fluence of $1 \times 10^{16}$ Au$^{4+}$ cm$^{-2}$. The mean dose in the foil is taken at the approximate middle of the foil (40 nm). The mean doses in LaMnO$_3$ and SrTiO$_3$ are 54.9 and 44.6 dpa, as shown in Figures \ref{si_srim_lmo} and \ref{si_srim_sto}, respectively. The maximum Au concentration retained in the foil at the depth of 80 nm is 0.3 at. \% in LaMnO$_3$ and 0.2 at. \% in SrTiO$_3$.

\clearpage

\section*{Supporting Information Note 4: STEM-EELS Measurements of Intermixing}

\begin{figure*}[h]
\includegraphics[width=0.5\textwidth]{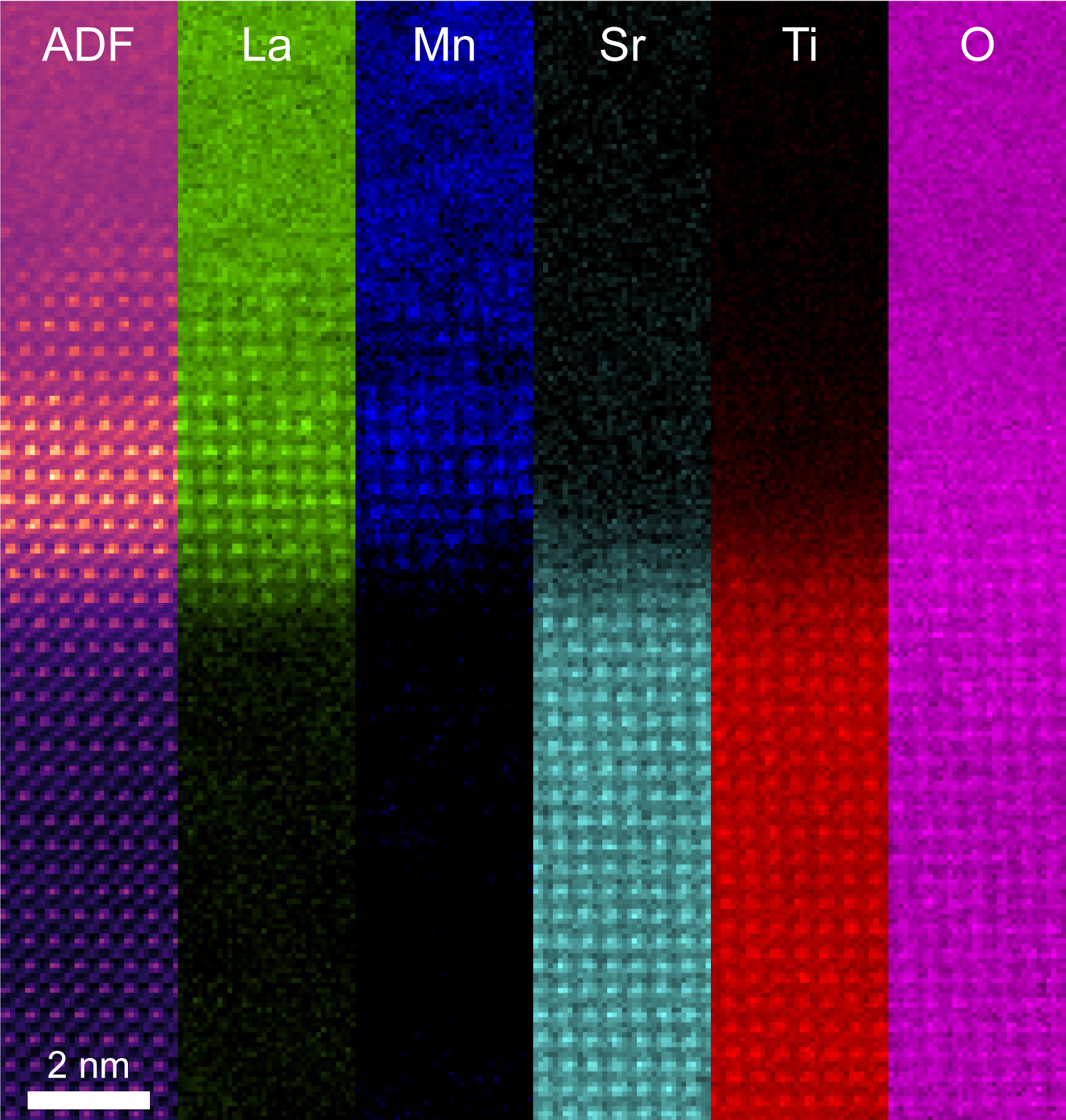}
\caption{Colorized STEM-ADF and corresponding STEM-EELS maps for the La $M_{4,5}$, Mn $L_{2,3}$, Sr $L_{2,3}$, and Ti $L_{2,3}$ edges. \label{si_eels_intermixing}}
\end{figure*}

STEM-EELS measurements were conducted after irradiation to assess potential post-irradiation intermixing. As shown in Figure \ref{si_eels_intermixing}, there is a non-negligible amount of cation intermixing on both the $A$- and $B$-sites, raising the possibility that this may contribute to the enhanced irradiation stability of the film at the interface.

\clearpage

\section*{Supporting Information Note 5: Irradiation Sequence}

The following images from main text Figure 1 are presented in greater detail to show the lattice structure and associated details during irradiation.

\begin{figure*}[h]
\includegraphics[width=\textwidth]{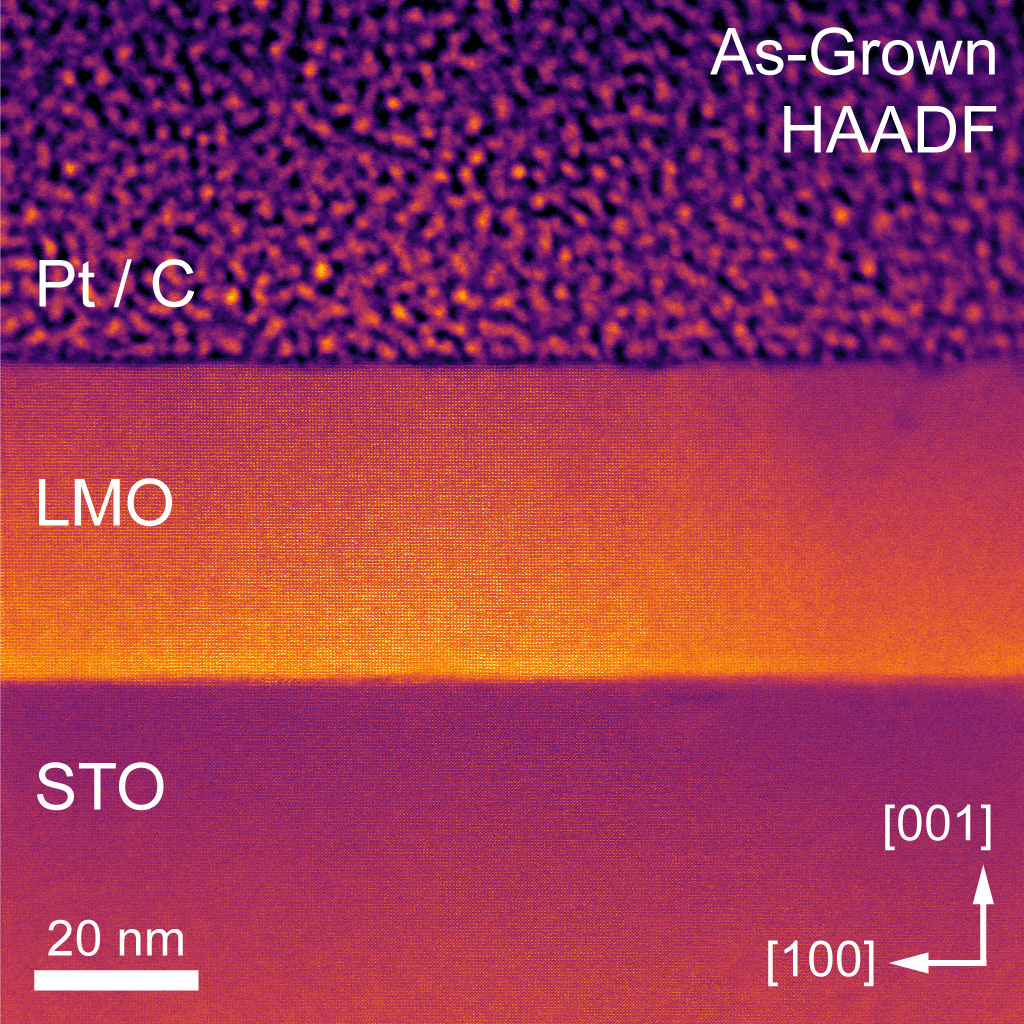}
\caption{Evolution of local disorder with fluence. Colorized cross-sectional STEM-HAADF image of the as-grown film.}
\end{figure*}

\begin{figure*}[h]
\includegraphics[width=\textwidth]{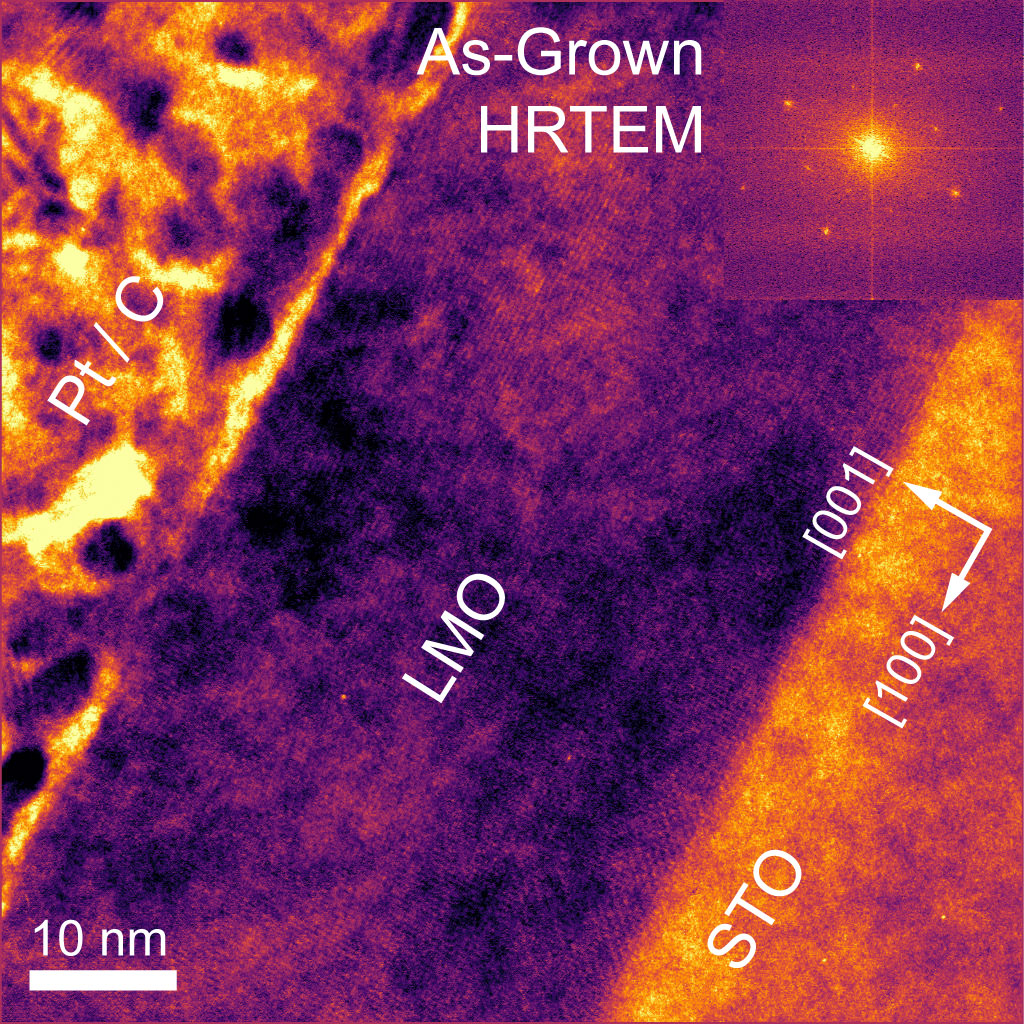}
\caption{Evolution of local disorder with fluence. Colorized cross-sectional HRTEM image of the as-grown film.}
\end{figure*}

\begin{figure*}[h]
\includegraphics[width=\textwidth]{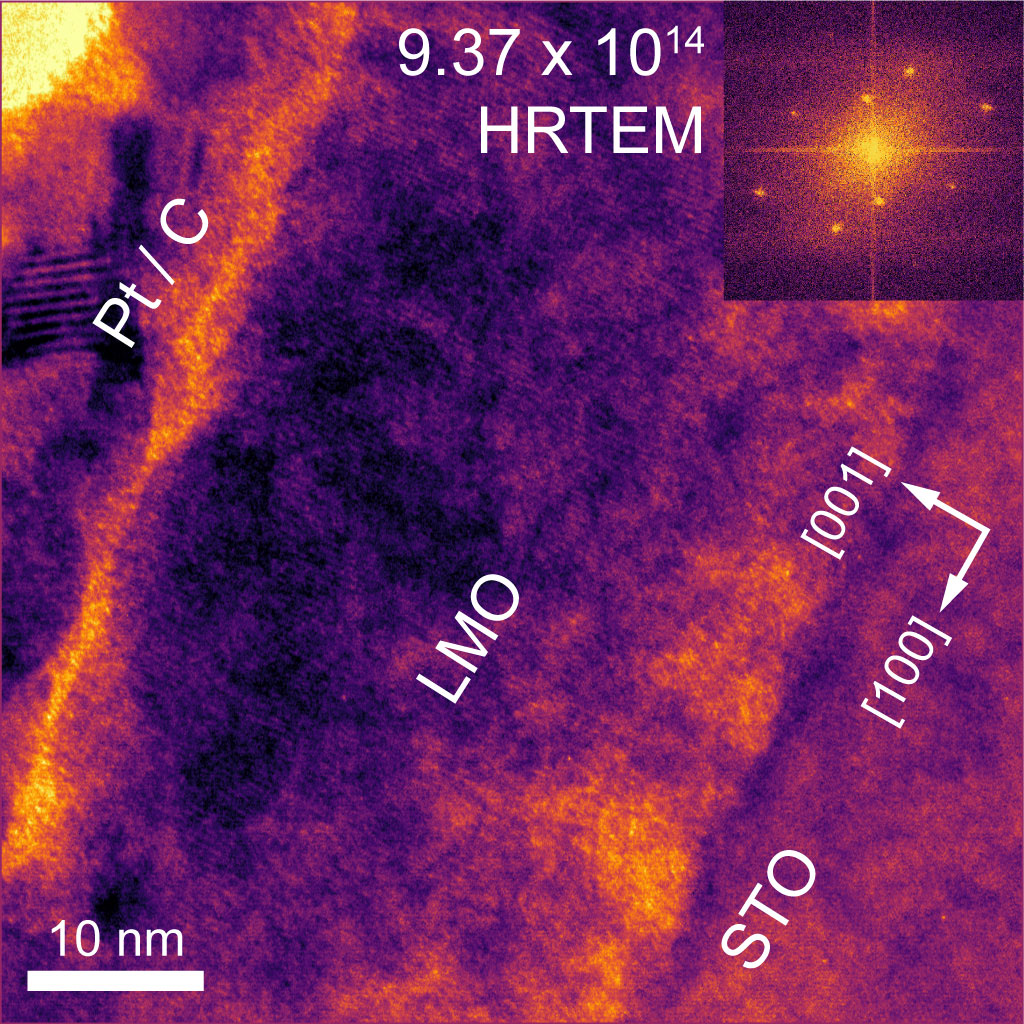}
\caption{Evolution of local disorder with fluence. Colorized cross-sectional HRTEM image of the $9.37 \times 10^{14}$ Au$^{4+}$ cm$^{-2}$ irradiated film.}
\end{figure*}

\begin{figure*}[h]
\includegraphics[width=\textwidth]{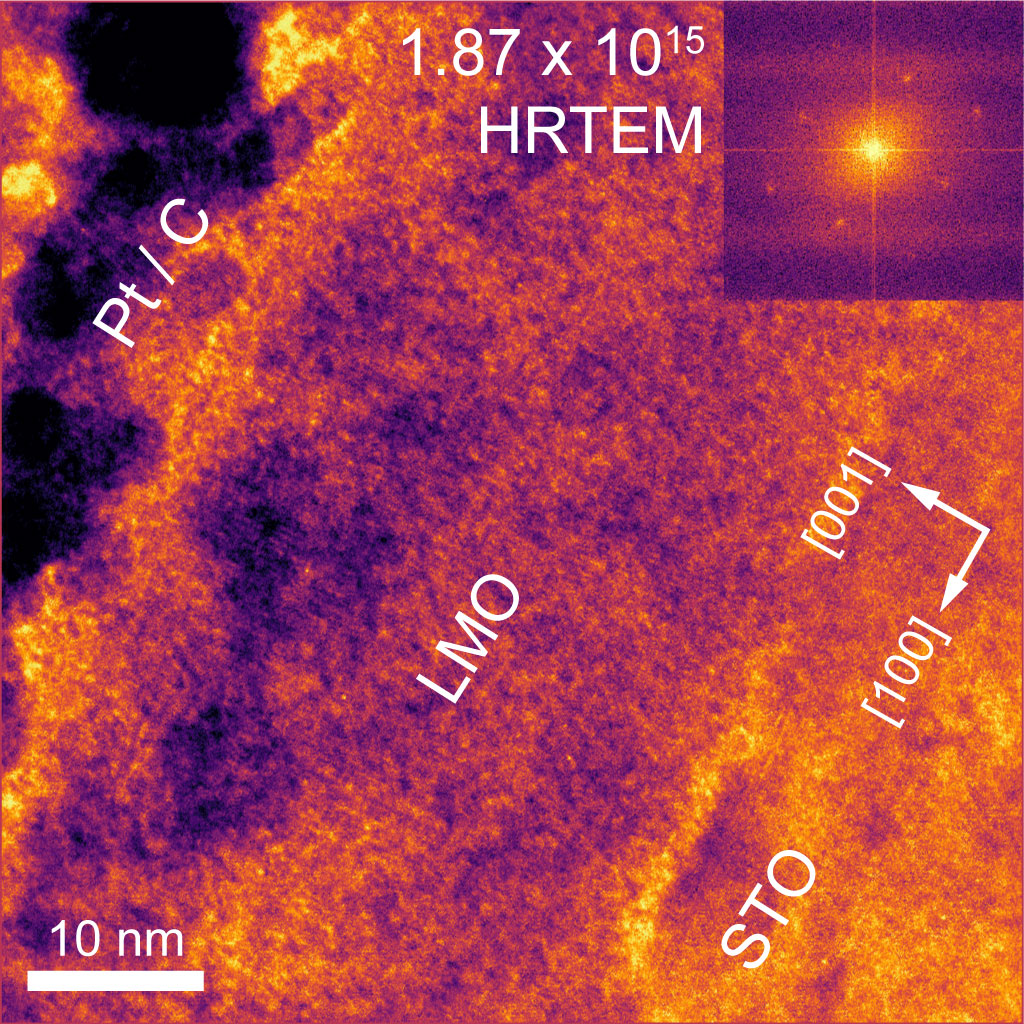}
\caption{Evolution of local disorder with fluence. Colorized cross-sectional HRTEM image of the $1.87 \times 10^{15}$ Au$^{4+}$ cm$^{-2}$ irradiated film.}
\end{figure*}

\begin{figure*}[h]
\includegraphics[width=\textwidth]{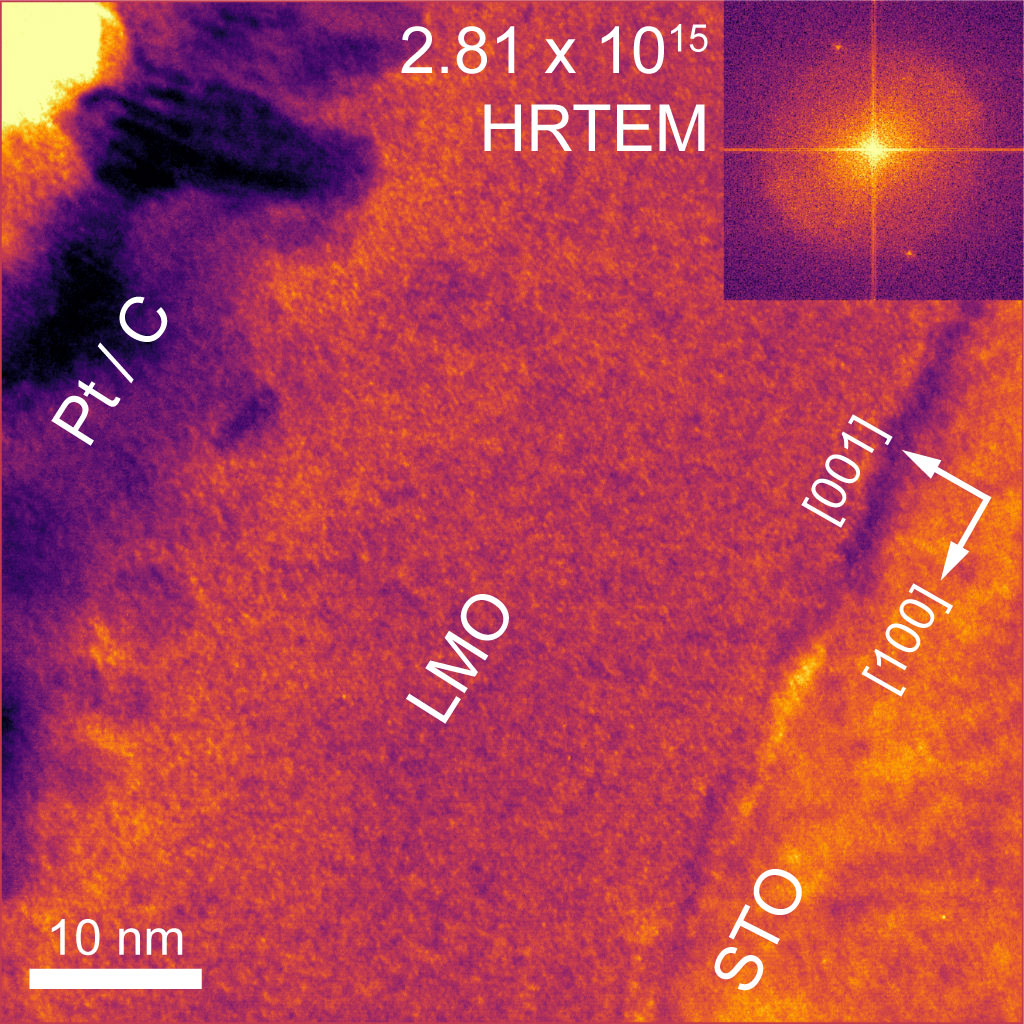}
\caption{Evolution of local disorder with fluence. Colorized cross-sectional HRTEM image of the $2.81 \times 10^{15}$ Au$^{4+}$ cm$^{-2}$ irradiated film.}
\end{figure*}

\begin{figure*}[h]
\includegraphics[width=\textwidth]{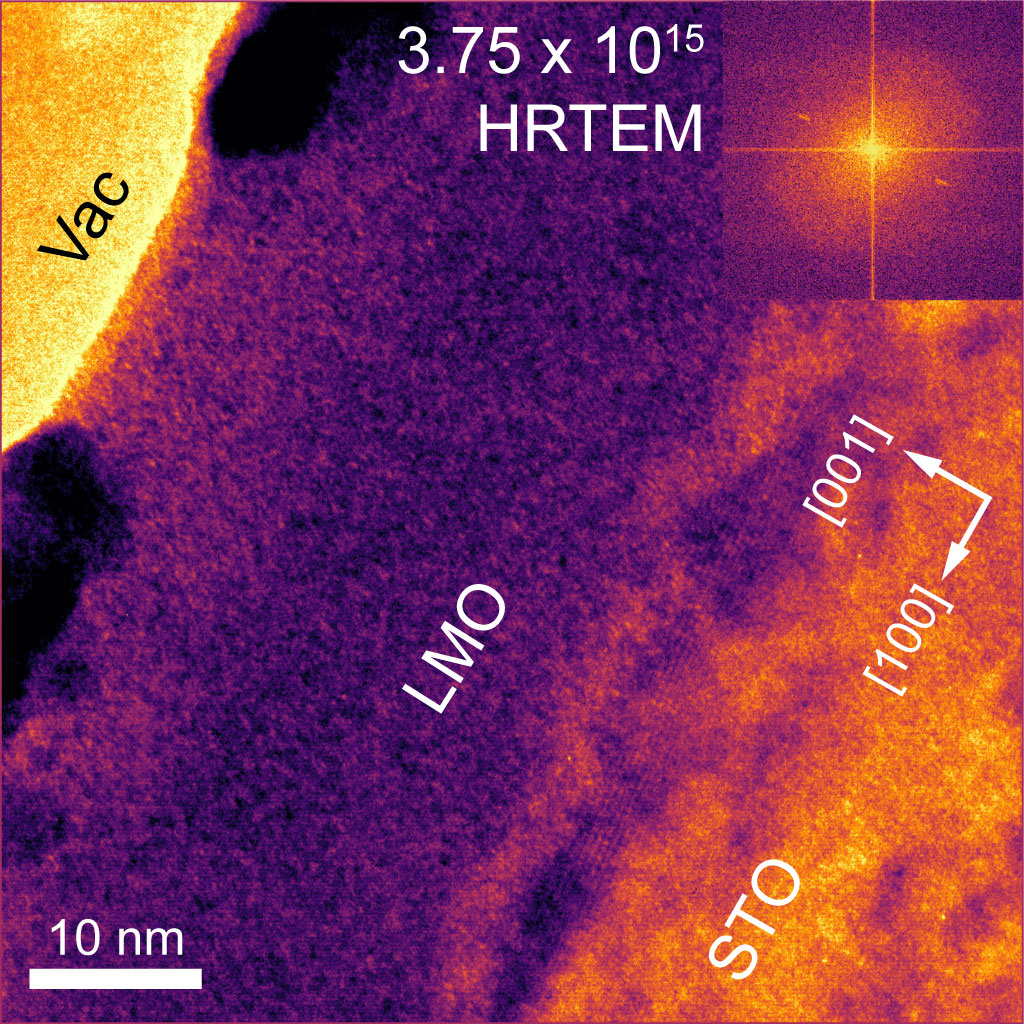}
\caption{Evolution of local disorder with fluence. Colorized cross-sectional HRTEM image of the $3.7 \times 10^{15}$ Au$^{4+}$ cm$^{-2}$ irradiated film.}
\end{figure*}

\begin{figure*}[h]
\includegraphics[width=\textwidth]{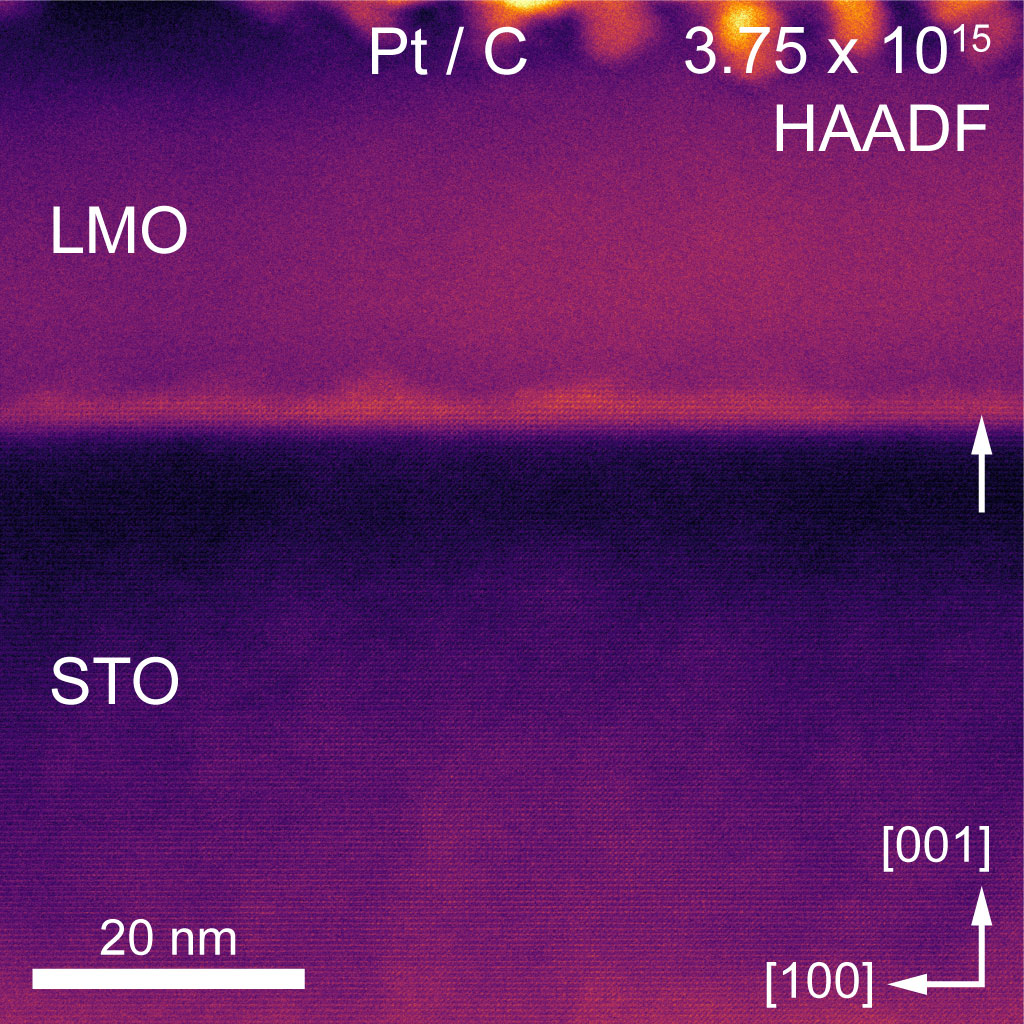}
\caption{Evolution of local disorder with fluence. Colorized cross-sectional STEM-HAADF of the $3.7 \times 10^{15}$ Au$^{4+}$ cm$^{-2}$ irradiated film.}
\end{figure*}

\clearpage

\section*{Supporting Information Note 6: \textit{In Situ} Movie}

Supporting Movie 1 was created by processing the \textit{in situ} movie using Matlab codes. First, the movie dataset from frames 1 to 8400 were Fourier downsampled to $400 \times 400$ pixels, such that the primary lattice planes were just below the Nyquist sampling frequency. Next, a Tukey edge window was applied to each image to create smooth boundary conditions, and the relative translation between images was measured using cross correlation. All images were then zero-padded to an image size of $800 \times 800$ pixels, and the measured translations were applied to align all images. Finally, these images were Bragg filtered from one of the primary lattice diffraction spots, using a moving average of 11 frames to reduce flickering of the outputs. The output movie images were constructed in a hue-saturation-value ($HSV$) color space. The $H$ channel was set to the lattice displacement (complex angle of the Bragg filter output) with periodic wrapping of output colors. The $S$ channel was set to a scaled Bragg filter magnitude (absolute value of the Bragg filter output), such that the image is grayscale where no lattice planes are present, and fully color-saturated where the lattice plane signals are strongest. Finally the $V$ channel is set to the original image intensity, scaled for best visibility. The Fourier transform amplitude of each image is inset into the upper right corner.

\clearpage

\bibliography{references}